\documentclass[12pt]{article}
\newcommand{\beq}{\begin{equation}}
\newcommand{\eeq}{\end{equation}}
\newcommand{\ba}{\begin{array}}
\newcommand{\ea}{\end{array}}
\newcommand{\beqa}{\begin{eqnarray}}
\newcommand{\eeqa}{\end{eqnarray}}
\newcommand{\lsim}{\stackrel{<}{_\sim}}

\newcommand{\epoe}{\epsilon'/\epsilon}

\newcommand{\tit}{{\tilde t}}
\newcommand{\tiu}{{\tilde u}}

\newcommand{\cA}{{\cal A}}
\newcommand{\cO}{{\cal O}}
\newcommand{\dis}{\displaystyle}

\newcommand{\PL}[3]{{\it Phys.\ Lett.\ }        {\bf #1} {#3} {(19#2)}}

\newcommand{\PRL}[3]{{\it Phys.\ Rev.\ Lett.\ } {\bf #1} {#3} {(19#2)}}
\newcommand{\PR}[3]{{\it Phys.\ Rev.\ }         {\bf #1} {#3} {(19#2)}}
\newcommand{\NP}[3]{{\it Nucl.\ Phys.\ }        {\bf #1} {#3} {(19#2)}}
\newcommand{\NC}[3]{{\it Nuovo\  Cimento\ }     {\bf #1} {#3} {(19#2)}}
\newcommand{\ZP}[3]{{\it Z.\ Phys.\ }           {\bf #1} {#3} {(19#2)}}
\newcommand{\IJMP}[3]{{\it Int.\ J.\ Mod.\ Phys.\ }{\bf #1} {#3} {(19#2)}}
\newcommand{\PTP}[3]{{\it Prog.\ Theor.\ Phys.\ }  {\bf #1} {#3} {(19#2)}}
\newcommand{\EPJ}[3]{{\it Eur.\ Phys.\ J.\ }       {\bf #1} {#3} {(19#2)}}
\newcommand{\JHEP}[3]{{\it JHEP\ }              {\bf #1} {#3} {(19#2)}}

\title{Standard Model vs  New Physics\\ 
in Rare Kaon Decays\thanks{Work  supported in part
by the EEC-TMR Program, Contract N.~CT98-0169.}}

\author{ 
\setcounter{footnote}{6}
Gino Isidori\thanks{INFN, Laboratori Nazionali 
di Frascati, P.O. Box 13, 00044
Frascati (Rome), Italy. } }
\date{}

\begin{document}
\maketitle

\begin{abstract}
\noindent
We present a brief overview of rare $K$ decays, emphasizing 
the different role of Standard Model and possible New Physics 
contributions in various channels.
\end{abstract}

\noindent
Being sensitive to flavour dynamics from few MeV up to 
several TeV, rare kaon decays provide a powerful tool 
to test the Standard Model (SM) and to search for New
Physics (NP). In the following we shall outline 
some of the most interesting aspects of these decays,
starting with the most rare ones, strongly sensitive 
to NP effects, moving toward processes which 
are more and more dominated by low-energy dynamics.

\section{Lepton-flavour violating modes}
Decays like $K_L\to \mu e$ and $K\to \pi \mu e$
are completely forbidden within the SM, where
lepton-flavour is conserved, but are also absolutely 
negligible if we simply extend the model by 
including only Dirac-type neutrino masses. 
A positive evidence of any of these processes 
would therefore unambiguously signal NP, calling 
for non-minimal extensions of the SM.
Moreover, as long as the final state contains 
at most one pion in addition to the lepton pair, 
the experimental information 
on the decay rate can be easily translated 
into a precise information on the 
short-distance amplitude $s \to d \mu e$.
In this respect we stress that $K_L\to \mu e$ and 
$K\to \pi \mu e$ provide a complementary 
information: the first mode is sensitive to 
pseudoscalar and axial-vector $s \to d$
couplings, whereas the second one is sensitive 
to scalar, vector an tensor structures. 

In exotic scenarios, like $R$-parity violating SUSY 
or models with leptoquarks, the $s \to d \mu e$ amplitude 
can be generated already at tree level. In this case naive 
power counting suggests that limits on $B(K_L\to \mu e)$ or 
$B(K\to \pi \mu e)$ at the level of $10^{-11}$ 
probe NP scales of the order of 100 TeV \cite{Peccei}.
On the other hand, in more ``conservative'' scenarios 
where the $s \to d \mu e$ transition can occur only at
the one-loop level, it is more appropriate 
saying that the scale probed is around the 
(still remarkable !) value of 1 TeV. 
An interesting example of the second type of 
scenarios is provided by left-right models 
with heavy Majorana neutrinos \cite{Aposto}.

\section{$K\to\pi\nu\bar{\nu}$} 
These decays are particularly fascinating since on one side,
within the SM, their small but non negligible rates are 
calculable with high accuracy in terms of the less known 
Cabibbo-Kobayashi-Maskawa (CKM) angles \cite{CKM}. 
On the other side, the flavour-changing 
neutral-current (FCNC) nature implies a strong 
sensitivity to possible NP contributions, even at very 
high energy scales.

Within the SM the $s \to d \nu \bar{\nu}$ amplitude
is generated only at the quantum level, through $Z$--penguin 
and $W$--box diagrams. 
Separating the contributions to the amplitude according to the 
intermediate up-type quark running inside the loop, one can write
\beq 
\cA(s \to d \nu \bar{\nu}) = \sum_{q=u,c,t} V_{qs}^*V_{qd} \cA_q 
\sim \left\{ \begin{array}{ll} \cO(\lambda^5
m_t^2)+i\cO(\lambda^5 m_t^2)\    & (q=t) \\
\cO(\lambda m_c^2 )\ + i\cO(\lambda^5 m_c^2)     & (q=c) \\
\cO(\lambda \Lambda^2_{QCD})    & (q=u)
\end{array} \right. \!
\label{uno}
\eeq
where $V_{ij}$ denote the elements of the CKM matrix. 
The hierarchy of these elements \cite{Wolf} 
would favor  up- and charm-quark contributions,
however the hard GIM mechanism of the parton-level calculation
implies $\cA_q \sim m^2_q/M_W^2$, leading to a completely 
different scenario. As shown on the r.h.s. of (\ref{uno}), 
where we have employed the standard phase convention 
($\Im V_{us}=\Im V_{ud}=0$) and 
expanded the CKM matrix in powers of the 
Cabibbo angle ($\lambda=0.22$) \cite{Wolf},
the top-quark contribution dominates both real and
imaginary parts.\footnote{~The $\Lambda^2_{QCD}$ factor in the last
line of (\protect\ref{uno})
follows from a naive estimate of long-distance effects.}
This structure implies several interesting consequences for
$\cA(s \to d \nu \bar{\nu})$: it is dominated by short-distance
dynamics and therefore calculable with high precision in perturbation
theory; it is very sensitive to $V_{td}$, which is one of the less
constrained CKM matrix elements;
it is likely to have a large $CP$-violating phase; it is very
suppressed within the SM and thus very sensitive to possible NP effects.

The short-distance contributions to $\cA(s \to d \nu \bar{\nu})$,
within the SM, can be efficiently described 
by means of a single effective dimension-6 operator:
$O^{\nu}_{LL}= (\bar{s}_L\gamma^\mu d_L)(\bar{\nu}_L \gamma_\mu \nu_L)$. The
Wilson coefficient of this operator has been calculated by Buchalla and
Buras including next-to-leading-order QCD corrections \cite{BB} (see
also \cite{MU,BB2}), leading to a very precise description of the partonic
amplitude.
Moreover, the simple structure of $O^{\nu}_{LL}$ has two major
advantages: 
\begin{itemize}
\item{} the relation between partonic and hadronic amplitudes 
is quite accurate, since the hadronic matrix elements
of the $\bar{s} \gamma^\mu d$ current between a kaon and a pion
are related by isospin symmetry to those entering $K_{l3}$ 
decays, which are experimentally well known; 
\item{} the lepton pair is produced in a state of definite $CP$ 
and angular momentum, implying that the leading SM contribution 
to $K_L \to \pi^0  \nu \bar{\nu}$ is $CP$ violating.
\end{itemize}

\subsection{SM uncertainties}
The dominant theoretical error in estimating 
$B(K^+\to\pi^+ \nu\bar{\nu})$
is due to the uncertainty of the QCD
corrections to the charm contribution
(see \cite{BB2} for an updated discussion), which 
can be translated into a $5\%$ error in the determination
of $|V_{td}|$ from $B(K^+\to\pi^+ \nu\bar{\nu})$. 
This uncertainty can be considered
as generated by `intermediate-distance' dynamics; genuine long-distance
effects associated to the up quark 
have been shown to be substantially smaller
\cite{LW}.

The case of $K_L\to\pi^0 \nu\bar{\nu}$ is even more clean from the
theoretical point of view \cite{Litt}. Indeed, because of the  $CP$
structure, only the imaginary parts in (\ref{uno}) 
-where the charm contribution is absolutely negligible-
contribute to $\cA(K_2 \to\pi^0 \nu\bar{\nu})$. Thus 
the dominant direct-$CP$-violating component 
of $\cA(K_L \to\pi^0 \nu\bar{\nu})$ is completely saturated by 
the top contribution, where the QCD uncertainties are 
very small (around 1\%). 
Intermediate and long-distance effects in this process
are confined only to the indirect-$CP$-violating 
contribution \cite{BB3} and to the $CP$-conserving one 
\cite{CPC} which are both extremely small.
Taking into account also the isospin-breaking corrections to the hadronic
matrix element \cite{MP}, one can therefore write a very accurate
expression (with a theoretical error around $1\%$)
for $B(K_L\to\pi^0 \nu\bar{\nu})$ in terms of short-distance parameters
\cite{BB2,BB3}: \beq
B(K_L\to\pi^0 \nu\bar{\nu})_{\rm SM}~=~4.25 \times 10^{-10}~\left[
\frac{\overline{m}_t(m_t) }{ 170~{\rm GeV}} \right]^{2.3} ~\left[ \frac{\Im
\lambda_t }{ \lambda^5 } \right]^2~. \eeq

The high accuracy of the theoretical predictions of $B(K^+ \to\pi^+
\nu\bar{\nu})$ and $B(K_L \to\pi^0 \nu\bar{\nu})$ in terms of the modulus
and the imaginary part of $\lambda_t= V^*_{ts} V_{td}$ could clearly offer
the possibility of very interesting tests of the CKM mechanism. Indeed, a
measurement of both channels would provide two independent information on
the unitarity triangle, which can be probed also by $B$-physics
observables. 
In particular, as emphasized in \cite{BB3}, the ratio of the two branching
ratios could be translated into a clean and complementary 
determination of $\sin(2\beta)$.

Taking into account all the indirect constraints on $V_{ts}$ and $V_{td}$
obtained within the SM, the present range of the SM predictions for the two
branching ratios reads \cite{BB2}: \beqa
B(K^+ \to\pi^+ \nu\bar{\nu})_{\rm SM} &=& (0.82 \pm 0.32) \times 10^{-10}~,
\label{BRK+nnt}\\
B(K_L \to\pi^0 \nu\bar{\nu})_{\rm SM} &=& (3.1 \pm 1.3) \times 10^{-11}~.
\label{BRKLnnt}
\eeqa
Moreover, 
As pointed out recently in \cite{BB2}, a stringent and 
theoretically clean upper bound on $B(K^+ \to\pi^+ \nu\bar{\nu})_{\rm SM}$ 
can be obtained using only the experimental 
information on $\Delta M_{B_d}/\Delta M_{B_s}$ to constraint 
$|V_{td}/V_{ts}|$. In particular, using
$(\Delta M_{B_d}/\Delta M_{B_s})^{1/2} < 0.2$ it is found 
\beq
B(K^+ \to\pi^+ \nu\bar{\nu})_{\rm SM} < 1.67 \times 10^{-10}~,
\eeq
which represents a very interesting challenge for the 
BNL-E787 experiment \cite{nt}.

\subsection{Beyond the SM: general considerations}
As far as we are interested only in 
$K \to\pi \nu\bar{\nu}$ decays, we can roughly distinguish 
the extensions of the SM into two big groups: those
involving new sources of quark-flavour mixing 
(like generic SUSY extensions of the
SM, models with new generations of quarks, etc\ldots) and those 
where the quark mixing is still ruled by the CKM matrix 
(like the 2-Higgs-doublet model of type
II, constrained SUSY models, etc\ldots). In the second case 
NP contributions are typically smaller than SM ones 
at the amplitude level
(see e.g. \cite{WWZ,SUSYc} for some recent discussions). 
On the other hand, in the
first case it is possible to overcome the $\cO(\lambda^5)$
suppression of the dominant SM amplitude. If this is the 
case, it is then easy to generate sizable enhancements 
of $K \to \pi\nu\bar{\nu}$ rates (see e.g. \cite{CI} and
\cite{fourth}).

Concerning $K_L\to\pi^0\nu\bar\nu$, it is worthwhile to emphasize
that if lepton-flavor is not conserved \cite{GN,GP} or right-handed 
neutrinos are involved \cite{BN}, then new $CP$-conserving 
contributions could in principle arise.

Interestingly, despite the variety of NP models, it is possible to derive a
model-independent relation among the widths of the three neutrino modes
\cite{GN}. Indeed, the isospin structure of any $s\to d$ operator bilinear
in the quark fields implies
\beq
\Gamma(K^+\to\pi^+\nu\bar{\nu}) =
\Gamma(K_L\to\pi^0\nu\bar{\nu}) +
\Gamma(K_S\to\pi^0\nu\bar{\nu}) ~,
\label{Tri}
\eeq
up to small isospin-breaking corrections, which then leads to
\beq
B(K_L\to\pi^0\nu\bar{\nu})< \frac{\tau_{_{K_L}}}{\tau_{_{K^+}}}
B(K^+\to\pi^+\nu\bar{\nu})\simeq 4.2 B(K^+\to\pi^+\nu\bar{\nu})~.
\label{GNbd}
\eeq
Any experimental limit on $B(K_L\to\pi^0\nu\bar{\nu})$ below this bound can
be translated into a non-trivial dynamical information on the structure of
the $s\to d\nu\bar{\nu}$ amplitude.

\subsection{SUSY contributions and the $Z \bar{s} d$ vertex}
We will now discuss in more detail the possible modifications 
of $K\to\pi\nu\bar{\nu}$ decays in the framework of a generic 
low-energy supersymmetric extension of the SM, 
which represents a very attractive possibility from the theoretical
point of view \cite{Hall}. Similarly to the SM, also in this case 
FCNC amplitudes are generated only at the quantum level, provided we 
assume unbroken $R$ parity and minimal particle content.
However, in addition to the standard penguin and box diagrams, also their
corresponding superpartners,  generated by gaugino-squarks loops, play an
important role. In particular, the chargino-up-squarks diagrams provide
the potentially dominant non-SM effect to the $s \to d \nu \bar{\nu}$ 
amplitude \cite{MWBRS}. Moreover, in the limit
where the average mass of SUSY particles
is substantially larger than $M_W$,
the penguin diagrams tend to dominate over the box ones and the dominant
SUSY effect can be encoded through an effective
$Z \bar{s} d$ coupling \cite{CI,BCIRS}.

The flavour structure of a generic SUSY model is quite complicated
and a convenient model-independent parameterization of the
various flavour-mixing
terms is provided by the so-called mass-insertion approximation
\cite{HKR}. This consists of choosing a simple basis for the gauge 
interactions and, in that basis, to perform a 
perturbative expansion of the squark mass matrices
around their diagonal. Employing a 
squark basis where all quark-squark-gaugino vertices
involving down-type quarks are flavor diagonal, it is found 
that the potentially dominant SUSY contribution to the 
$Z \bar{s} d$ vertex arises from the double 
mixing $(\tiu^{d}_L - \tit_R) \times (\tit_R - \tiu^{s}_L)$ \cite{CI}. 
Indirect  bounds on these mixing terms  dictated 
by vacuum-stability, neutral-meson mixing 
and $b \to s \gamma$ leave open the possibility of 
large effects \cite{CI}. More stringent constraints 
can be obtained employing stronger theoretical assumptions 
on the flavour structure of the SUSY model \cite{BCIRS}.
However, the possibility of sizable modifications 
of $K\to\pi\nu\bar{\nu}$ widths 
(including enhancements of more than one order of magnitude 
in the case of  $K_L \to\pi^0 \nu\bar{\nu}$)
cannot be excluded a priori.

Interestingly a non-standard $Z \bar{s} d$ vertex can be 
generated also in non-SUSY extensions of the SM 
(see e.g. \cite{NS}). It is therefore useful trying to 
constraint this scenario in a model-independent way.
At present the best direct limits on 
the $Z \bar{s} d$  vertex are dictated by $K_L\to\mu^+\mu^-$
\cite{GN,CI,BS}, 
bounding the real part of the coupling, and $\Re(\epoe)$ \cite{BS},
constraining the imaginary one.
Unfortunately in both cases the bounds are not very 
accurate, being affected by sizable hadronic 
uncertainties. Concerning $\epoe$,
it is worthwhile to mention that the non-standard 
$Z \bar{s} d$ vertex could provide an explanation for
the apparent discrepancy between $(\epoe)_{\rm exp}$
and $(\epoe)_{\rm SM}$ \cite{BCIRS,sanda},
even if it is certainly too early to make definite 
statement in this respect \cite{epop}.
In the future the situation could become much 
more clear with precise determinations of both 
real and imaginary part of the $Z_{\bar{s} d}$ 
coupling by means of $\Gamma(K^+ \to\pi^+ \nu\bar{\nu})$
and $\Gamma(K_L \to\pi^0 \nu\bar{\nu})$.
Note that if we only use the present constraints from 
$K_L\to\mu^+\mu^-$ and $\Re(\epoe)$, we cannot exclude 
enhancements up to one order of magnitude for 
$\Gamma(K_L \to \pi^0 \nu\bar{\nu})$ and up to a factor 
$\sim 3$ for $\Gamma(K^+ \rightarrow \pi^+ \nu \bar\nu)$ 
\cite{BCIRS,BS}.

\section{$K\to\pi \ell^+\ell^-$ and  $K\to \ell^+\ell^-$}
Similarly to $K\to\pi\nu\bar{\nu}$ decays, 
the short-distance contributions to 
$K\to\pi \ell^+\ell^-$ and  $K\to \ell^+\ell^-$ 
are calculable with high accuracy and 
are potentially sensitive to NP effects.
However, in these processes the size of long-distance 
contributions is usually much larger due to 
the presence of electromagnetic interactions. 
Only in few cases (mainly in $CP$-violating observables)
long-distance contributions are suppressed and it 
is possible to extract the interesting short-distance
information. 

\subsection{$K\to \pi \ell^+\ell^-$} 
The single-photon exchange amplitude, dominated by long-distance dynamics,
provides the largest contribution to the $CP$-allowed transitions 
$K^+ \to \pi^+ \ell^+ \ell^-$ and $K_S \to \pi^0 \ell^+ \ell^-$.
The former has been observed, both in the electron and in the muon 
mode, whereas only an upper bound of about $10^{-6}$
exists on $B(K_S\to\pi^0 e^+e^-)$ \cite{PDG}.
This amplitude can be described in a model-independent way 
in terms of two form factors, $W_+(z)$ and $W_S(z)$,
defined by \cite{DEIP}
\beqa
& i \dis\int d^4x e^{iqx} \langle \pi(p)|T \left\{J^\mu_{\rm elm}(x)
{\cal L}_{\Delta S=1}(0) \right\} |
K_i (k)\rangle =& \nonumber \\
&\dis\frac{W_i(z)}{(4\pi)^2}\left[z(k+p)^\mu -(1-r_\pi^2)q^\mu
\right]~, & \label{eq:tff}
\eeqa
where $q=k-p$, $z=q^2/M_K^2$ and $r_\pi = M_\pi /M_K$.
The two form factors are non singular at $z=0$ and,
due to gauge invariance, vanish to lowest order in 
Chiral Perturbation Theory (CHPT) \cite{EPR}.
Beyond lowest order one can identify 
two separate contributions to the $W_i(z)$:
a non-local term, $W_i^{\pi\pi}(z)$, due to the 
$K\to 3\pi\to \pi\gamma^*$ scattering, 
and a local term, $W_i^{\rm pol}(z)$,
that encodes the contributions of unknown 
low-energy constants (to be determined by data) \cite{DEIP}.
At $\cO(p^4)$ the local term is simply a constant,
whereas at $\cO(p^6)$ also a term linear in $z$ arises.
We note, however, that already at  $\cO(p^4)$
chiral symmetry alone does not help to 
relate $W_S$ and $W_+$, 
or $K_S$ and $K^+$ decays \cite{EPR}.

Recent results on $K^+ \to \pi^+ e^+ e^-$ and
$K^+ \to \pi^+ \mu^+ \mu^-$ by  BNL-E865 \cite{E865}
indicates very clearly that, 
due to a large linear slope, the $\cO(p^4)$ 
expression of $W_+(z)$ is not sufficient
to describe experimental data.
This should not be consider as a failure of CHPT, 
rather as an indication that large $\cO(p^6)$ contributions
are present in this channel.\footnote{~This should not 
surprise since in this mode sizable next-to-leading
order contributions could arise due to vector-meson
exchange.} Indeed the $\cO(p^6)$ expression of  
$W_+(z)$ seems to fit well data. Interestingly, 
this is not only due to a new free parameter
appearing at $\cO(p^6)$, but it is also due to the 
presence of the non-local term. The evidence 
of the latter provides a real significant test 
of the CHPT approach.

In the $K_L \to \pi^0 \ell^+ \ell^-$ decay the long-distance part of the
single-photon exchange amplitude is forbidden by $CP$ invariance but it
contributes to the processes via $K_L$-$K_S$ mixing, leading to
\beq
B(K_L \to \pi^0 e^+ e^-)_{\rm CPV-ind}~=~ 3\times 10^{-3}~ B(K_S \to \pi^0 e^+
e^-)~.
\eeq
On the other hand, the direct-$CP$-violating part of the decay amplitude
is very similar to the one of $K_L \to \pi^0 \nu \bar{\nu}$ but for the
fact that it receives an additional short-distance contribution due 
to the photon penguin. Within the SM, this theoretically clean part 
of the amplitude leads to
\cite{BLMM}
\beq
B(K_L\to\pi^0 e^+e^-)^{\rm SM}_{\rm CPV-dir}~=~0.67 \times 10^{-10}~\left[
\frac{\overline{m}_t(m_t) }{ 170~{\rm GeV}} \right]^{2} ~\left[ \frac{\Im
\lambda_t }{ \lambda^5 } \right]^2~,
\eeq
and, similarly to the case of $B(K_L \to \pi^0 \nu \bar{\nu})$, it 
could be substantially enhanced by SUSY contributions \cite{CI,BCIRS}.
The two $CP$-violating components of the $K_L\to\pi^0 e^+e^-$ amplitude
will in general interfere. 
Given the present uncertainty on $B(K_S \to
\pi^0 e^+ e^-)$, at the moment we can only set the rough upper limit
\beq
B(K_L\to\pi^0 e^+e^-)_{\rm CPV-tot}^{\rm SM}~\lsim~{\rm few}\times 10^{-11}
\label{BRKLet}
\eeq
on the sum of all the $CP$-violating contributions to this mode 
\cite{DEIP}. We stress, however, that the phases of the two $CP$-violating
amplitudes are well know. Thus if $B(K_S \to \pi^0 e^+ e^-)$ will be
measured, it will be possible to determine the interference between direct
and indirect $CP$-violating components of $B(K_L\to\pi^0 e^+e^-)_{\rm
CPV}$ up to a sign ambiguity. Finally, it is worth to note that
an evidence of $B(K_L\to\pi^0 e^+e^-)_{\rm CPV}$ above the $10^{-10}$ level,
possible within specific supersymmetric scenarios \cite{BCIRS}, 
would be a clear signal of physics beyond the SM.

An additional contribution to $K_L \to \pi^0 \ell^+ \ell^-$ 
decays is generated by the $CP$-conserving processes
$K_L \to \pi^0 \gamma \gamma \to \pi^0 \ell^+ \ell^-$ \cite{Sehgal}. 
This however does not interfere with the 
$CP$-violating amplitude and, as we shall discuss in the next
section, it is quite small ($\lsim~4\times 10^{-12}$)
in the case of $K_L \to \pi^0 e^+e^-$.

\subsection{$K_L \to l^+ l^-$}
The two-photon intermediate state plays  
an important role in $K_L\to\ell^+\ell^-$ transitions.
This is by far the dominant contribution
in $K_L\to e^+ e^-$, where the dispersive integral 
of the $K_L \to \gamma \gamma \to l^+ l^-$ loop
is dominated by the term proportional to $\log(m_K^2/m^2_e)$.
The presence of this large logarithm implies also that 
$\Gamma(K_L\to e^+ e^-)$ can be estimated with a relatively 
good accuracy in terms of $\Gamma(K_L\to \gamma \gamma)$,
yielding to the prediction
$B(K_L \to e^+ e^-) \sim 9 \times 10^{-12}$ \cite{VP} 
which recently seems
to have been confirmed by the four events observed at
BNL-E871 \cite{E871}.

More interesting from the short-distance point of view is the case of $K_L
\to \mu^+\mu^-$. Here the two-photon long-distance amplitude is 
not enhanced by large logs and it is almost comparable in size
with the short-distance one, sensitive to $\Re V_{td}$ \cite{BB}. 
Unfortunately the dispersive part of the two-photon
contribution is much more difficult to be estimated in this case, due to
the stronger sensitivity to the $K_L \to \gamma^* \gamma^*$ form factor.
Despite the precise experimental determination of $B(K_L \to \mu^+\mu^-)$,
the present constraints on $\Re V_{td}$ from this observable are not very
stringent \cite{DIP}. Nonetheless, the measurement of $B(K_L \to
\mu^+\mu^-)$ is still useful to put significant bounds on possible NP
contributions. Moreover, we stress that the uncertainty of the $K_L \to
\gamma^*\gamma^*\to \mu^+\mu^-$ amplitude could be partially decreased in
the future by precise experimental information on the form factors of $K_L
\to \gamma \ell^+\ell^-$ and
$K_L \to e^+e^- \mu^+\mu^-$ decays, especially if these would be consistent
with the general 
parameterization of the $K_L \to \gamma^* \gamma^*$ vertex
proposed in \cite{DIP}.

\section{Two-photon processes} 
$K\to \pi \gamma\gamma$ and $K\to \gamma\gamma$ decays 
are completely dominated by short distance dynamics and 
therefore not particularly useful to search for NP.
However, these modes are interesting on one side to 
perform precision 
tests of CHPT, on the other side 
to estimate long-distance corrections to the $\ell^+\ell^-$ channels
(see e.g. \cite{review} and references therein).

Among the CHPT tests, an important role is 
played by  $K_S \to \gamma \gamma$.
The first non-vanishing contribution to this process 
arises at $\cO(p^4)$ and, 
being generated only by a finite loop amplitude,
is completely determined \cite{Espriu}.
Since in this channel vector meson exchange contributions 
are not allowed, and unitarity corrections are 
automatically included in the $\cO(p^2)$ coupling \cite{review},
we expect that  
the $\cO(p^4)$ result provides a good approximation 
to the full amplitude. This is confirmed by present data \cite{PDG},
but a more precise determination of the branching ratio 
is need in order to perform a more stringent test.

Similarly to the  $K_S \to \gamma \gamma$ case, also 
the leading non-vanishing contribution to
$K_L \to \pi^0 \gamma \gamma$  arises 
only at $\cO(p^4)$ and is completely determined \cite{KLp0gg}.
However, in this case large $\cO(p^6)$
corrections can be expected due to both unitarity corrections 
and vector meson exchange contributions.
Indeed the $\cO(p^4)$ prediction for 
$B(K_L\to\pi^0\gamma\gamma)$ turns out to be 
substantially smaller (more than a factor 2) 
than the experimental findings \cite{review}.
After the inclusion of unitarity corrections
and vector meson exchange contributions, 
both spectrum and branching ratio of 
this decay can be expressed in terms 
of a single unknown coupling: $a_V$ 
\cite{KLggee}. 
The recent KTeV measurement \cite{KTeV} 
has shown that the determination of $a_V$ 
from both spectrum and branching ratio of
$K_L\to\pi^0\gamma\gamma$ leads to the same value,
$a_V = -0.72 \pm 0.08$, providing an important
consistency check of this approach.

As anticipated, the  $K_L \to \pi^0 \gamma \gamma$ amplitude  
is also interesting since it produces a
$CP$-conserving contribution to $K_L \to\pi^0 \ell^+ \ell^-$ 
\cite{KLggee}. For $\ell=e$ the leading $O(p^4)$
contribution is helicity suppressed and only 
the $O(p^6)$ amplitude with the two photons in $J=2$ 
leads to a non-vanishing $B(K_L \to\pi^0 e^+ e^-)_{\rm CPC}$ 
\cite{Sehgal}. Given the recent experimental result \cite{KTeV}, 
this should not exceed $4\times 10^{-12}$ \cite{KLggee}. 
Moreover, we stress that the Dalitz
plot distribution of $CP$-conserving  and $CP$-violating 
contributions to $K_L \to\pi^0 e^+
e^-$ are substantially different: in the first case the $e^+e^-$ pair is
in a state of $J=1$, whereas in the latter has $J=2$. Thus in 
principle it is possible to extract the interesting
$B(K_L \to\pi^0 e^+ e^-)_{\rm CPV}$ from a Dalitz plot analysis
of the decay. On the other hand, the
$CP$-conserving contribution is enhanced and more difficult 
to be subtracted in the case of $K_L \to\pi^0 \mu^+ \mu^-$,
where the helicity suppression of the 
leading $O(p^4)$ contribution (photons in $J=0$) 
is much less effective
(see Heiliger and Sehgal in \cite{KLggee}).

\section{Conclusions}
Rare $K$ decays  provide a unique opportunity to
perform high precision tests of $CP$ violation and flavour mixing,
both within and beyond the SM. 

A special role is undoubtedly played by $K \to \pi \nu\bar{\nu}$ decays.
In some NP scenarios sizable enhancements to 
the branching ratios of these modes are 
possible and, if detected, these would provide the 
first evidence for physics beyond the SM. 
Nevertheless, even in absence of such enhancements,
precise measurements of $K \to \pi \nu\bar{\nu}$ widths
will lead to unique information about the flavour 
structure of any extension of the SM. 

Among decays into a $\ell^+\ell^-$ pair, the most
interesting one from the short-distance point of view is
probably $K_L \to \pi^0 e^+ e^-$.  However, in order to extract 
precise information from this mode an experimental 
determination (or a stringent upper bound)
on $B(K_S \to\pi^0 e^+ e^-)$ is also necessary.

\end{document}